\documentclass[fleqn,usenatbib]{mnras}

\usepackage{newtxtext,newtxmath}

\usepackage[T1]{fontenc}
\usepackage{ae,aecompl}

\usepackage{graphicx}	
\usepackage{amsmath}	
\usepackage{amssymb}	

\title[MACS J0744.9+3927: absence of radio relic]{Search for low-frequency diffuse radio emission around a shock in the massive galaxy cluster MACS J0744.9+3927}
\author[A. Wilber et al.]{
A. Wilber$^{1}$\thanks{E-mail: amanda.wilber@hs.uni-hamburg.de (A. Wilber)},
M. Br{\"u}ggen$^{1}$,
A. Bonafede$^{1,2}$,
D. Rafferty$^{1}$,
F. Savini$^{1}$,
\newauthor
T. Shimwell$^{3}$,
R. J. van Weeren$^{4}$,
A. Botteon$^{5,2}$,
R. Cassano$^{2}$,
G. Brunetti$^{2}$,
\newauthor
F. De Gasperin$^{4}$,
D. Wittor$^{1}$,
M. Hoeft$^{6}$,
L. Birzan$^{1}$
\\
$^{1}$Hamburger Sternwarte, University of Hamburg, Gojenbergsweg 112, 21029 Hamburg, Germany\\
$^{2}$INAF/Istituto di Radioastronomia, Via P Gobetti 101, 40129 Bologna, Italy\\
$^{3}$ASTRON, the Netherlands Institute for Radio Astronomy, Postbus 2, 7990 AA, Dwingeloo, The Netherlands\\
$^{4}$Leiden Observatory, Leiden University, PO Box 9513, 2300 RA Leiden, The Netherlands\\
$^{5}$Dipartimento di Fisica e Astronomia, Università di Bologna, via P. Gobetti 93/2, 40129 Bologna, Italy\\
$^{6}$Th\"uringer Landessternwarte, Tautenburg, Germany\\ 
}

\date{Accepted XXX. Received YYY; in original form ZZZ}

\pubyear{2015}

\begin{document}
\label{firstpage}
\pagerange{\pageref{firstpage}--\pageref{lastpage}}
\maketitle
%
\begin{abstract}
Merging galaxy clusters produce low Mach number shocks in the intracluster medium. These shocks can accelerate electrons to relativistic energies that are detectable at radio frequencies. MACS J0744.9+3927 is a massive ($M_{500} = (11.8 \pm 2.8) \times 10^{14} M_{\odot}$), high-redshift ($z=0.6976$) cluster where a Bullet-type merger is presumed to have taken place. Sunyaev-Zel'dovich maps from MUSTANG indicate that a shock, with Mach number $\mathcal{M} = 1.0-2.9$ and an extension of $\sim 200$ kpc, sits near the centre of the cluster. The shock is also detected as a brightness and temperature discontinuity in X-ray observations. To search for diffuse radio emission associated with the merger, we have imaged the cluster with the LOw Frequency ARray (LOFAR) at 120-165 MHz. Our LOFAR radio images reveal previously undetected AGN emission, but do not show clear cluster-scale diffuse emission in the form of a radio relic nor a radio halo. The region of the shock is on the western edge of AGN lobe emission from the brightest cluster galaxy. Correlating the flux of known shock-induced radio relics versus their size, we find that the radio emission overlapping the shocked region in MACS J0744.9+3927 is likely of AGN origin. We argue against the presence of a relic caused by diffusive shock acceleration and suggest that the shock is too weak to accelerate electrons from the intracluster medium.

\end{abstract}




\section{Introduction}

\subsection{Cluster mergers and shocks}

Mergers between galaxy clusters produce large-scale, low-Mach number ($\mathcal{M} \leq 4 - 5$) shock waves in the intracluster medium (ICM). These shocks are thought to accelerate ICM particles to ultra-relativistic energies and potentially amplify ICM magnetic fields \citep[e.g.][]{2009MNRAS.395.1333V, 2012SSRv..166..187B, 2017MNRAS.464..210V}. Observations of merging clusters over a wide range of wavelengths are currently constraining the physics behind these shocks. See \citet{2007MNRAS.375...77H} and \citet{2014IJMPD..2330007B} for reviews of shocks and non-thermal emission associated with these events.\\

The thermal component of the ICM consists of hot gas ($10^{7-8}$ K or 1-10 keV) that is visible in X-rays through bremstrahlung emission. A merging cluster may show an elongated or disturbed X-ray morphology, usually indicating the direction of the merger. Shocks are pressure discontinuities that can be identified as sharp edges in the brightness and temperature distribution of the ICM X-ray emission \citep[e.g.][]{2002ApJ...567L..27M, 2007PhR...443....1M, 2013MNRAS.429.2617O, 2017arXiv170707038B}. \\

The mass and merging status of a cluster can also be inferred from the thermal Sunyaev-Zel'dovich (SZ) effect, where electrons in the ICM up-scatter cosmic microwave background (CMB) photons.  The SZ decrement\footnote{A decrement is seen only below $\sim$~220 GHz; above this frequency there is an increment.} is proportional to the line-of-sight integral of the plasma pressure and hence shocks appear as substructures in the SZ signal of a cluster. Since the SZ decrement has no dependence on the distance, there is the opportunity to discover shocks even in distant clusters \citep{2001PASJ...53...57K, 2004PASJ...56...17K, 2011ApJ...734...10K, 2010ApJ...716..739M, 2011ApJ...734...10K, 2015ApJ...807..121R, 2015ApJ...809..185Y}. \citet{2011A&A...534L..12F} were the first to use MUSTANG SZ maps and X-ray data to confirm a shocked region in the most X-ray luminous cluster RX J1347-1145, and showed that there was a radio excess coincident with the shock in the form of a mini halo. Recently the SZ effect was used to characterize a shock in the Coma cluster using Planck data \citep{2015MNRAS.447.2497E}. More recently, ALMA has achieved very high resolutions (up to 3.5 arcsec), detecting the highest redshift shock known ($z=0.87$) in ACT-CL J0102-4915 or the `El Gordo' cluster \citep{2016ApJ...829L..23B}.\\

Many merging cluster systems are observed to host cluster-scale radio emission in the form of radio halos and/or radio relics \citep[see][for a review]{feretti2012}. Radio halos are classified as diffuse emission at the cluster centre, thought to be the product of turbulent re-acceleration of ICM particles driven by a cluster merger \citep[e.g.][]{2001MNRAS.320..365B}. Gischt relics are classified as elongated or arching diffuse radio emission typically found on the cluster periphery \citep[e.g.][]{1997MNRAS.290..577R, 2006Sci...314..791B, 2009A&A...494..429B, 2010Sci...330..347V, 2012A&A...546A.124V, 2014ApJ...785....1B}, and are thought to trace shock waves induced by cluster mergers \citep{2011MmSAI..82..627B, 2012SSRv..166..187B}. Active galactic nuclei (AGN) relics, or phoenix relics, are usually much smaller, more roundish in appearance, and consist of aged or ``ghost" AGN emission that has been re-energized by shocks \citep[e.g][]{2001A&A...366...26E, 2001AJ....122.1172S, 2015MNRAS.448.2197D}. \\

The details of the necessary acceleration mechanisms and the efficiencies of low Mach number shocks are still largely unknown \citep[see][for a review]{2014IJMPD..2330007B}. Along a shock front, diffusive shock acceleration (DSA) is believed to accelerate electrons that could produce synchrotron emission at radio wavelengths \citep{1978ApJ...221L..29B, 1998A&A...332..395E}.
However, the low Mach numbers of most observed shocks are usually too weak to accelerate particles from the thermal pool \citep{2016ApJ...818..204V, 2016MNRAS.460L..84B, 2016MNRAS.461.1302E}. Hence, some form of pre-acceleration or an upstream population of relativistic seed electrons is required for DSA to operate. One possible pre-acceleration mechanism is shock drift acceleration (SDA) \citep{2011ApJ...742...47M, 2014ApJ...797...47G, 2014ApJ...794..153G}, which has been simulated in a cosmological context by \citet{2016MNRAS.459...70V} and  \citet{2017MNRAS.464.4448W}. AGN are another potential source for seed electrons, and several examples have been found, the clearest being the connection between a radio relic and radio galaxy in Abell 3411-3412 \citep{2017NatAs...1E...5V}.\\

Most merging clusters with X-ray detected shocks are shown to host some type of diffuse radio emission. There are a few cases where shocks coincide with the edges of radio halos \citep[e.g][]{2005ApJ...627..733M, 2016PASJ...68S..20U}, but most shocks in merging clusters are associated with radio relics. Very recently, \citet{2017arXiv170803641H} presented highly sensitive Jansky Very Large Array (JVLA) observations of the low mass merging cluster Abell 2146 that revealed diffuse radio emission associated with two confirmed X-ray shocks which was previously undetected by \citet{2011MNRAS.417L...1R} in observations by the Giant Meterwave Radio Telescope (GMRT). \\

Searching for diffuse radio emission at confirmed shock locations in the ICM is a key test to validate the widely held model that radio relics are produced by shocks. Radio observations of smaller shocks and shocks with lower Mach numbers are especially of interest, potentially yielding limits on the efficiencies of theorized acceleration mechanisms. High redshift clusters also test the sensitivities of our radio telescopes, and allow us to determine the effect of inverse Compton scattering which becomes stronger as the CMB energy density increases.

\subsection{MACS J0744.9+3927}

An intracluster shock sits just outside the centre of the massive and distant galaxy cluster MACS J0744.9+3927 \citep{2011ApJ...734...10K}. This cluster is located at RA: 07h44m52.47s, Dec: +39$^{\circ}$27\arcmin27.3\arcsec, at a redshift of $z=0.6976$ \citep{2007ApJ...661L..33E}. \citet{2016A&A...588A..69D} give a mass derived from the XMM Newton archival data as $M_{500} = 9.9\times10^{14}M_{\odot}$. MACS J0744.9+3927 is a CLASH (Cluster Lensing And Supernova
survey with Hubble; \citealp{2012ApJS..199...25P}) cluster with a weak lensing derived mass of $M_{500} = (11.837 \pm 2.786 )\times10^{14}M_{\odot}$ \citep{2015MNRAS.449.2024S}, consistent (within 1$\sigma$) with the XMM-derived mass.\\ 

Using MUSTANG observations, \citet{2011ApJ...734...10K} found an SZ-decrement as a kidney-shaped ridge in the north-south direction with a length of $\sim 25$ arcsec (180 kpc). The kidney-shaped feature is located between the system's main mass peak and a second, smaller mass peak. The region of this ridge overlaps a discontinuity in the Chandra X-ray image. To accurately quantify the discontinuity, \citet{2011ApJ...734...10K} measured X-ray surface brightness and temperature changes over the elliptical radius of the cluster, and found that there is temperature increase and a slight brightness drop-off coincident with the SZ decrement, suggesting that this region is shock-heated gas and the second highest redshift shock known after the one in El Gordo \citep{2016MNRAS.463.1534B}. \\

The Mach number obtained from the density jump conditions, as calculated from the fit to the X-ray surface brightness, was $\mathcal{M} = 1.2_{-0.2}^{+0.2}$ where the errors are $1\sigma$. The temperature jump conditions at the shock yield a higher value, $\mathcal{M} = 2.1^{+0.8}_{-0.5}$; however, the error bars are large and the value agrees with the Mach number inferred from the density jump condition at the $1.3\sigma$ level. The shock velocity in this cluster is $V_{\rm sh}=1827^{+267}_{-195}$ km s$^{-1}$ assuming the Mach number obtained from the density jump conditions \citep{2011ApJ...734...10K}.\\

\begin{figure}
\centering
\includegraphics[width=0.49\textwidth]{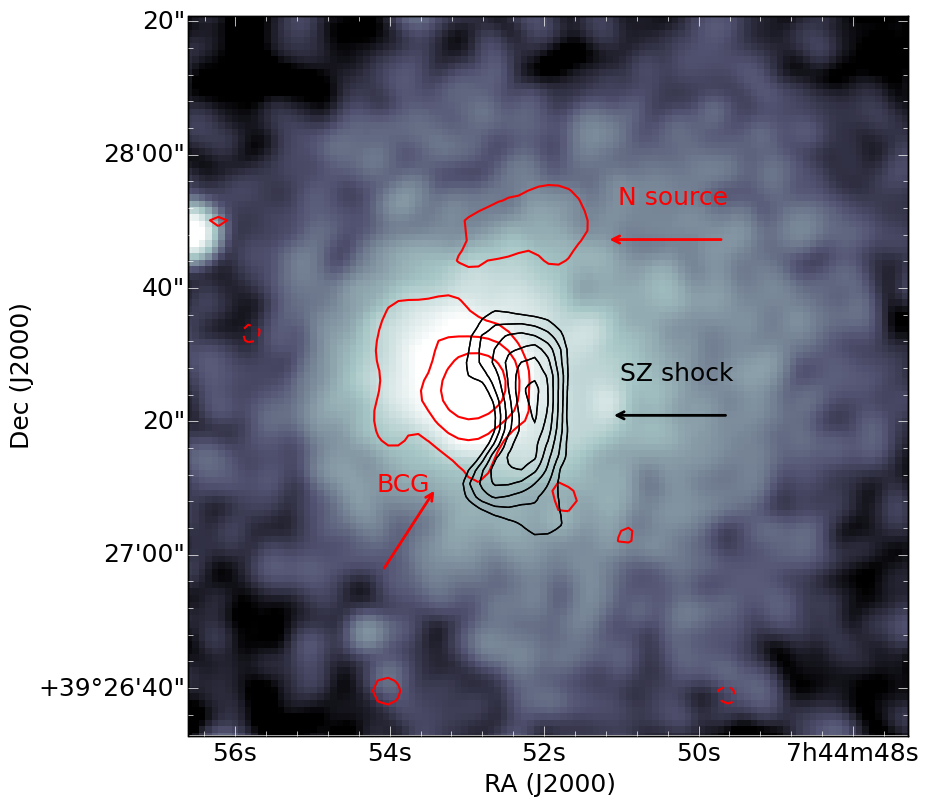}
\caption{Chandra X-ray emission in blue with LOFAR high-resolution FACTOR image contours $[-3, 3, 6, 12]~\times~\sigma$ overlaid in red. X-ray emission is smoothed with a Gaussian kernel. RMS noise of our LOFAR image is $\sigma= 180~\mu$Jy beam$^{-1}$ with a beam size of 8.6 arcsec $\times$ 6.5 arcsec. Black contours show the SZ-decrement as detected by MUSTANG, from \citealp{2011ApJ...734...10K}, starting at 3$\sigma$ with 0.5$\sigma$ increments. LOFAR detects a compact BCG AGN and a compact northern source.\label{factor}}
\end{figure}

The dynamics of the cluster-sub-cluster merger is unclear. A subcluster lies 300 kpc west of the centre of the main cluster and a mass lensing reconstruction from \citet{2011MNRAS.413..643R} shows elongation toward the west. \citet{2011ApJ...734...10K} suggest that the subcluster has passed through the main cluster core from the east to the west in a Bullet-type fashion and ram-pressure has stripped off baryons from the subcluster. Shocked gas appears to be hugging the westward portion of the main cluster core, in between the two merging clusters, which is not the place where a merger shock would be expected. Typically, the shocks found in Bullet-type mergers, either bow or counter, are found outside of the merged system as a whole, and not in between the two subclusters \citep[e.g][]{2010MNRAS.406.1721R, 2014MNRAS.440.2901S, 2016MNRAS.463.1534B}. The size and position of the shock suggest that it is not the type of merger shock that has been observed to produce prominent radio relics such as the Sausage \citep[e.g.][]{2017MNRAS.471.1107H} or Toothbrush \citep[e.g.][]{2016ApJ...818..204V} relics because those are all on the peripheral regions of the cluster and exhibit significant scale ($\sim1$ Mpc). \citet{2014A&A...561A.112G} presented an X-ray image of the cluster from XMM Newton and state that the residual X-ray image shows an extended structure coincident with the SZ excess as seen by \citet{2011ApJ...734...10K}, but they favour the scenario that the extended appearance is due to a substructure attached to the cluster, potentially from an infalling group. The direction of the potential infalling group and the angular separation between it and the main cluster is uncertain.\\ 

Radio surveys covering MACS J0744.9+3927, including the Faint Images of the Radio Sky at Twenty-Centimeters (FIRST; \citealt{1997ApJ...475..479W}) and the NRAO VLA Sky Survey (NVSS; \citealt{1998AJ....115.1693C}), do not detect any significant radio sources in the cluster field. MACS J0744.9+3927 was selected to be targeted with the LOw Frequency ARray (LOFAR; \citealp{vanHaarlem2013}) because it is a very high-mass and high-$z$ cluster with evidence of merging activity and a shock detection in SZ. In this paper we present LOFAR observations of MACS J0744.9+3927 at 120-165 MHz (with a central frequency of 143 MHz) to search for a radio relic associated with the shock. LOFAR's sensitivity to steep-spectrum low-surface-brightness emission is crucial for detecting weak and diffuse radio emission on cluster scales. Assuming the following cosmology, H$_{0}=70$, $\Omega_{m}=0.3$, and $\Omega_{\Lambda}=0.7$, the angular scale at MACS J0744.9+3927's redshift ($z=0.6976$) is 7.223 kpc arcsec$^{-1}$, used hereafter.

\section{LOFAR observation of MACS J0744.9+3927}

LOFAR is a low-frequency radio interferometer with a compact core and stations that extend over large parts of Northern Europe \citep{vanHaarlem2013}. Our observation was part of the LOFAR Two-meter Sky Survey (LoTSS; \citealt{2017A&A...598A.104S}) and uses the high-band antennas (HBA) over a frequency range of 120-165 MHz. The data reduction steps for this data are identical to the data reduction steps described in \citet{2017arXiv170808928W}, and are summarised below.  

\subsection{Prefactor}

Prefactor\footnote{\url{https://github.com/lofar-astron/prefactor}} is a package containing automated pipelines called Pre-Facet-Calibration and Initial-Subtract. Pre-Facet-Calibration compresses and averages the original data and performs all initial, direction-independent calibration. In this step a flux calibrator (observed at the beginning and end of the target observation) is used to compute amplitude gain solutions, station clock offsets, station phase offsets, and station differential total electron content (dTEC). Amplitude gain solutions and corrections for clock and phase offsets are then transferred to the target field data. An initial phase calibration is also performed using a global sky model from the TIFR GMRT Sky Survey (TGSS) at 150 MHz \citep{2017A&A...598A..78I}. For this observation the calibrator 3C196 was used, a bright quasar (66 Jy at 159 MHz according to the \citealt{2012MNRAS.423L..30S} absolute flux scale). After the direction-independent calibration is completed, preliminary imaging is carried out via the Initial-Subtract pipeline. The full wide-field of the calibrated target data is imaged in high and low resolution using {\tt WSClean} \citep{2014MNRAS.444..606O}. These full field images are used to model and subtract all sources in preparation for direction-dependent calibration. 

\subsection{FACTOR \label{Facet Calibration}}
Direction-dependent calibration for LoTSS data is significantly simplified and refined through the facet calibration technique \citep{vanWeeren16}. This method of calibration is executed via the FACTOR\footnote{\url{http://www.astron.nl/citt/facet-doc/}} software package. FACTOR first tesselates the full target field into several smaller patches of sky called facets. Each facet is automatically chosen to be centered on a bright compact source to be used as a facet calibrator. TEC, phase, and amplitude solutions are computed from the facet calibrator and applied to all sources in that facet. Facets are processed in order of brightness and as the brighter sources are progressively subtracted, with adequate calibration solutions, the effective noise in the $uv$-data is reduced. After a facet is calibrated, it is imaged with {\tt WSClean} and a primary beam correction is applied. For more details on facet calibration the reader is referred to \citet{vanWeeren16}, \citet{shimwell2016}, and \citet{2016MNRAS.460.2385W}.\\

\begin{figure}
\centering
\includegraphics[width=0.49\textwidth]{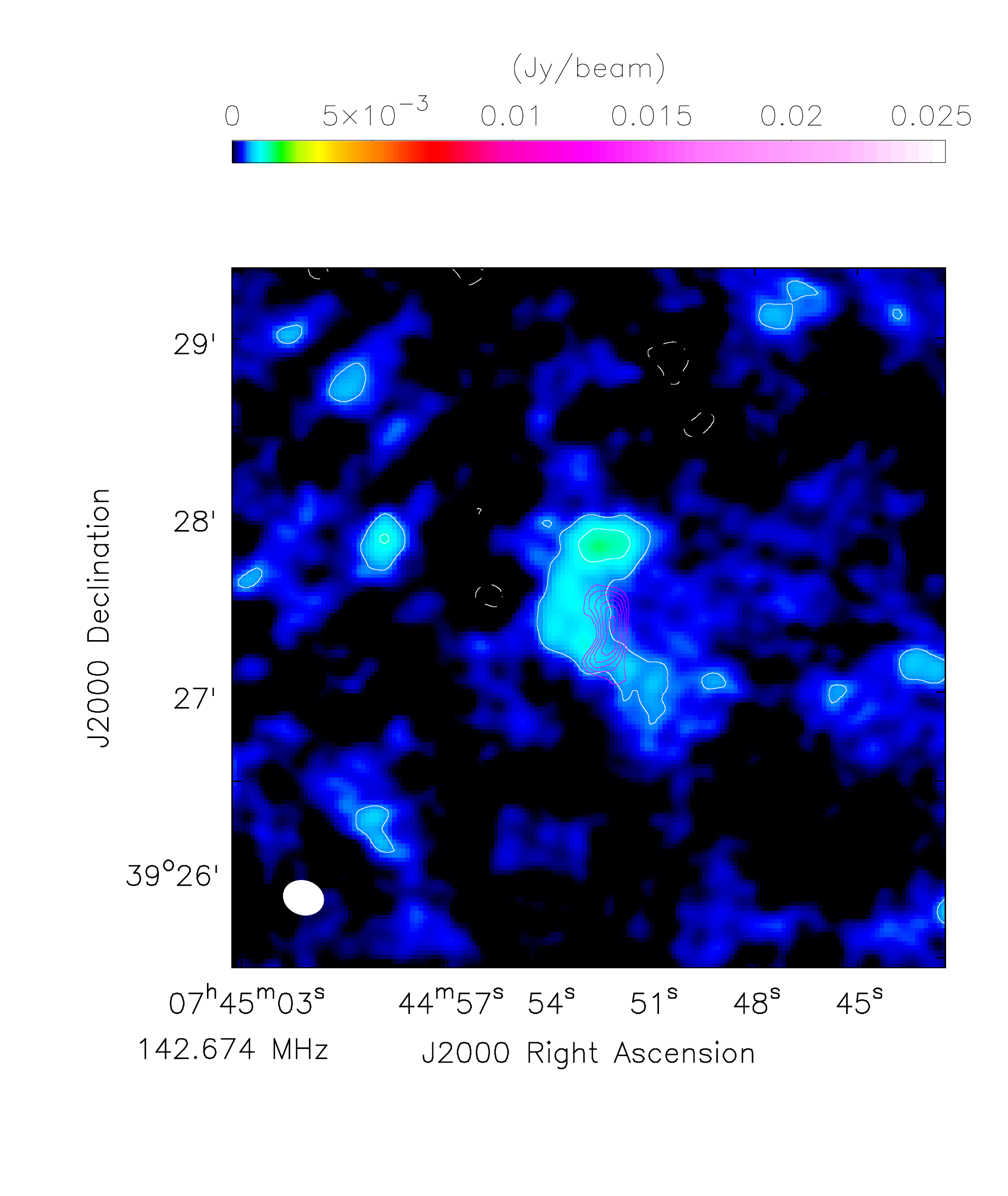}
\caption{LOFAR image after subtracting compact sources imaged at a uvrange of $>$ 6000 $\lambda$. An outer $uv$taper of 6 arcsec was used to bring out diffuse emission. RMS noise is $\sigma= 180~\mu$Jy beam$^{-1}$ and the restoring beam is 14 arcsec $\times$ 12 arcsec. White contours are $[-3,3,6,9]~\times~\sigma$. Magenta contours show the SZ feature, considered to be a merger-induced shock, as detected by MUSTANG from \citealp{2011ApJ...734...10K}. \label{t6}}
\end{figure}

\begin{figure}
\centering
\includegraphics[width=0.49\textwidth]{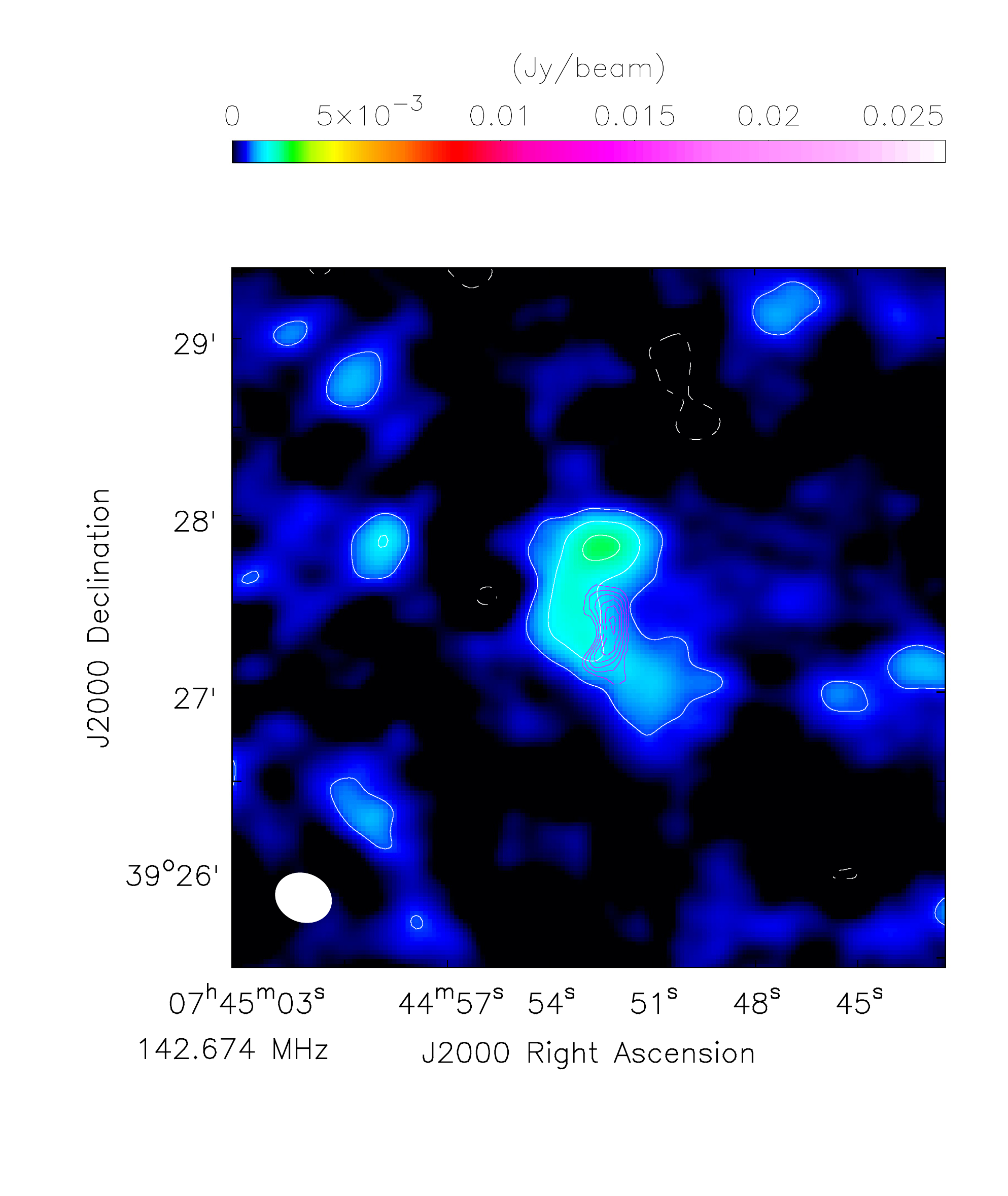}
\caption{LOFAR image after subtracting compact sources imaged at a uvrange of $>$ 6000 $\lambda$. An outer $uv$taper of 10 arcsec was used to bring out diffuse emission. White contours are $[-3,3,6,9]~\times~\sigma$ where $\sigma= 200~\mu$Jy beam$^{-1}$ and the restoring beam is 19 arcsec $\times$ 16 arcsec. Magenta contours are from the MUSTANG SZ map. Extended emission is visible up to 570 kpc, and likely consists of emission from multiple AGN. \label{t10}}
\end{figure}

For this observation, FACTOR was run on the full 45 MHz bandwidth of target data. The wide-field was tessellated into 40 directions with one direction designated as the target facet, containing MACS J0744.9+3927, and 18 bright and nearby facets were processed before imaging the target facet. The target facet was imaged after applying the calibration solutions from a nearby\footnote{$\sim 0.3^{\circ}$ angular separation} bright source in a neighboring facet. The final image of the target facet produced by FACTOR has a resolution of 8.6 arcsec $\times$ 6.5 arcsec with root mean square (rms) noise of $\sigma\approx180~\mu$Jy beam$^{-1}$. \\

Data calibrated with FACTOR often show a slight astrometric offset because phase solutions change quickly over small regions of sky (due to the spatially changing conditions of the ionosphere; \citealt{2016MNRAS.460.2385W}). The LOFAR map was initially offset from the optical and SZ map. We calculated an astrometric shift by comparing the LOFAR radio map at 143 MHz to a high resolution GMRT map at 610 MHz with corrected astrometry from van Weeren et al. (in preparation). We measured the offset in arcseconds between the maximum pixel locations of several point sources in and around the cluster centre. A shift of RA, Dec: [-2.5,0] arcsec was applied to all LOFAR images to match the GMRT astrometry.

\section{Results}

\subsection{Subtraction of compact sources}
Our LOFAR image of MACS J0744.9+3927 shows an active galaxy at the cluster centre. It sits within the X-ray emission of the cluster but slightly toward the east (see Fig.~\ref{factor}). This central AGN is likely associated with the optical source SDSS J074452.77+392726.7, the brightest cluster galaxy (BCG), with a redshift of $z=0.6986 \pm 0.0007$\footnote{from the Sloan Digital Sky Survey Data Release 1 as obtained Aug. 28, 2003 from \url{http://das.sdss.org/DR1/data/spectro/ss_tar_20/}}. The radio emission from this galaxy appears round and compact in the high-resolution image. There is another compact radio source on the northern edge of the X-ray emission (labeled in Fig.~\ref{factor}). This emission might come from an active galaxy, but the only optically visible galaxy coincident with the radio peak in this region does not have a confirmed redshift (SDSS J074452.36+392748.8). Therefore, it is not possible to say whether this northern emission is a background or foreground galaxy, or if it is actually extended lobe emission from the AGN associated with the BCG.\\

In order to search for cluster-scale diffuse emission associated with the shocked region, a subtraction of compact sources was carried out using CASA (Common Astronomy Software Applications; \citealp{2007ASPC..376..127M}). The subtraction was performed on the $uv$ data by imaging with a $uv$-range of $>$ 6000 $\lambda$ (filtering out emission that spans more than 34 arcsec or $\sim$ 250 kpc)\footnote{We also attempted the subtraction at a $uv$-range of $>$ 2000 $\lambda$ and $>$ 4000 $\lambda$ which correspond to emission spanning less than $\sim$ 750 kpc and $\sim$ 400 kpc, respectively, but decided to increase the cut to reduce the possibility of subtracting diffuse emission possibly related to the shocked region since this region is on a relatively small scale.} and Briggs' robust parameter of 0.25. The CLEAN components of the resulting image were then subtracted from the $uv$-dataset using the tasks FTW and UVSUB. The dataset was then re-imaged with a $uv$-range of $>$ 80 $\lambda$\footnote{FACTOR calibration solutions are determined from data with a $uv$-range of $>$ 80 $\lambda$ to eliminate the shortest baselines which can introduce significant large scale emission manifested as noise. Since the target data is calibrated with solutions $>$ 80 $\lambda$ we also image selecting data $>$ 80 $\lambda$.}, a slight outer $uv$-taper to bring out extended emission (6 arcsec as shown in Fig.~\ref{t6} and 10 arcsec as shown in Fig.~\ref{t10})\footnote{Outer $uv$-taper values greater that 10 arcsec appeared to artificially inflate the size of existing emission due to smearing from a larger synthesized beam. Since it is at such high redshift, this cluster requires higher-resolution imaging.}, and a robust parameter of 0. This method of compact source subtraction is not perfect and the resulting image may still include residual emission or artefacts associated with the compact source, or diffuse emission can be subtracted, but a manual inspection of the clean component model proved that no diffuse emission was removed. \\

\begin{figure}
\centering
\includegraphics[width=0.49\textwidth]{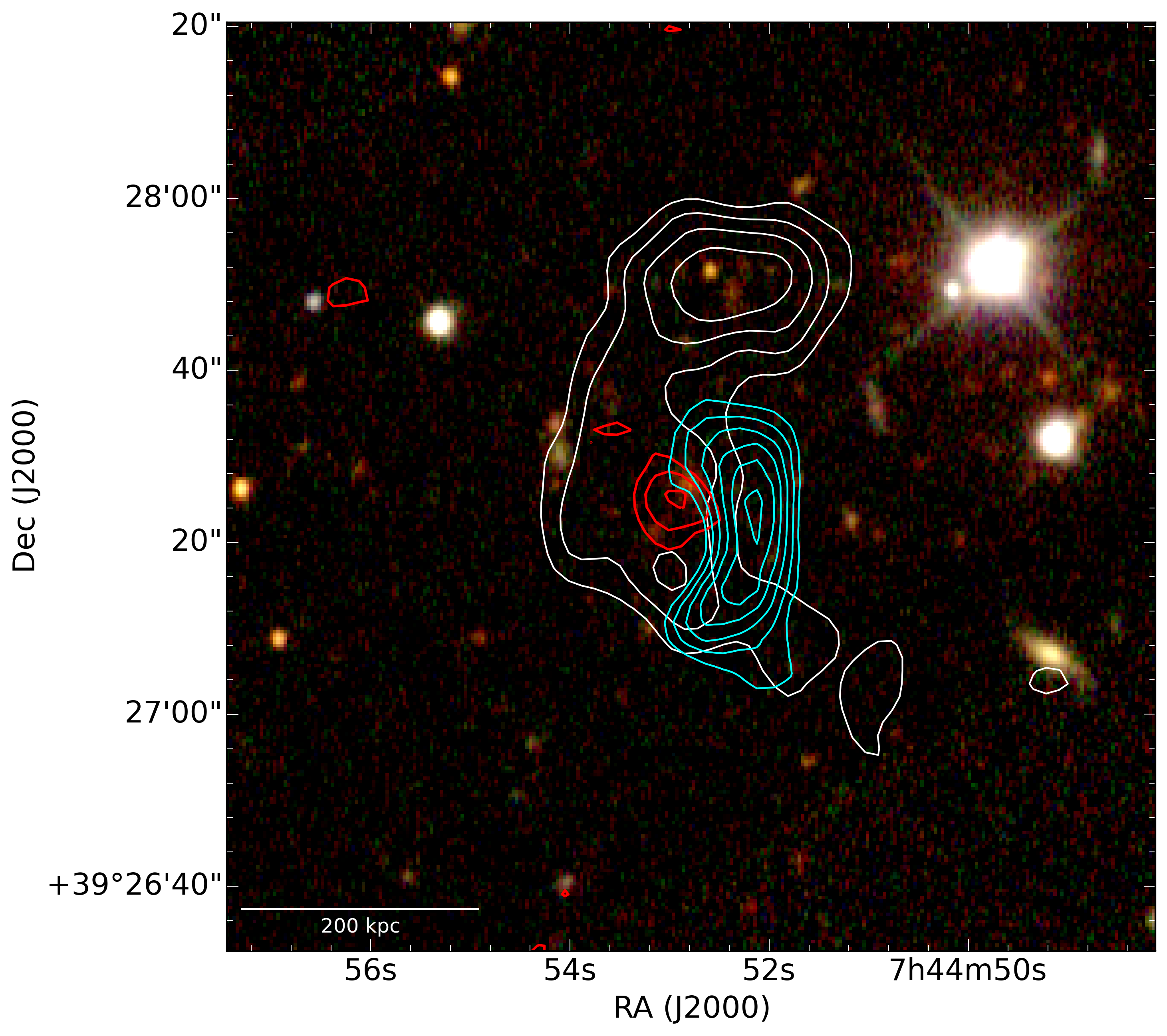}
\caption{SDSS {\it g,r,i} image with radio emission overlaid as contours. LOFAR compact emission imaged with a uvrange $>$ 6000 $\lambda$ is shown by red contours $[3,6,9]~\times~\sigma$ where $\sigma = 200\mu$Jy beam$^{-1}$. LOFAR diffuse emission after compact source subtraction is shown in white contours where contours are $[3, 4, 5, 6]~\times~\sigma$ and $\sigma=200\mu$Jy beam$^{-1}$ (same as Fig.\ref{t10}). Cyan contours are from the SZ MUSTANG map from \citealt{2011ApJ...734...10K}. \label{opt}}
\end{figure}

Our final compact-source-subtracted image shows faint diffuse emission near the cluster centre extending to the north and south of the BCG (Fig.~\ref{t6} \& Fig.~\ref{t10}). An optical image of the cluster with LOFAR compact emission and diffuse emission overlaid as contours can be seen in Fig.~\ref{opt}. The northern diffuse emission is brightest with a peak flux of 933 $\mu$Jy within the 9$\sigma$ contour where $\sigma = 200\, \mu$Jy beam$^{-1}$ (in Fig.~\ref{t10}), which is likely from a separate AGN north of the BCG. We suspect that the diffuse emission at the cluster centre extending southwest is part of radio lobes associated with the BCG. There is no significant emission coincident with the kidney-shaped ridge seen in the SZ map. In fact, the diffuse radio emission falls off in the direction that the SZ decrement increases (see Fig.~\ref{opt}). There is also no diffuse emission resembling a radio halo at the cluster centre.\\

\subsection{Search for radio relics}

\begin{table*}
 \centering
   \caption{Small \& faint radio relics / candidate radio relics} \label{obstable}
   \label{tab1}
  \begin{tabular}{c|c|c|c}
  \hline
Relic & log$_{10}$(P$_{1.4}$)[W Hz$^{-1}$] & LLS [Mpc] & References \\
\hline
A13 & 23.85 & 0.25  & \citet{feretti2012,2001AJ....122.1172S}\\
A85 & 23.50 & 0.35  & \citet{feretti2012,2001AJ....122.1172S}\\
A725 & 23.11 & 0.44  & \citet{2000ApJS..129..435B,2001ApJ...548..639K}\\
A2034 F &  22.60 & 0.6  & \citet{2016MNRAS.459..277S} \\
A2048 & 23.66 & 0.31  & \citet{vw11b} \\
A2443 & 23.30 & 0.43  & \citet{feretti2012,2011AJ....141..149C}\\
A4038 & 23.01 & 0.13  & \citet{feretti2012,2001AJ....122.1172S} \\
MAXBCG138+25 & 25.01 & 0.19  & \citet{vanW2011} \\
Sausage R1 & 24.03 & 0.63  & \citet{2017MNRAS.471.1107H}\\
Sausage R2 & 24.27 & 0.67  & \citet{2017MNRAS.471.1107H} \\
Toothbrush D & 24.15 & 0.25  & \citet{2016ApJ...818..204V} \\
24P73 & 23.88 & 0.27  & \citet{vw11b} \\
\hline
\hline
\end{tabular}
\end{table*}

\begin{figure}
\centering
\includegraphics[width=0.35\textwidth]{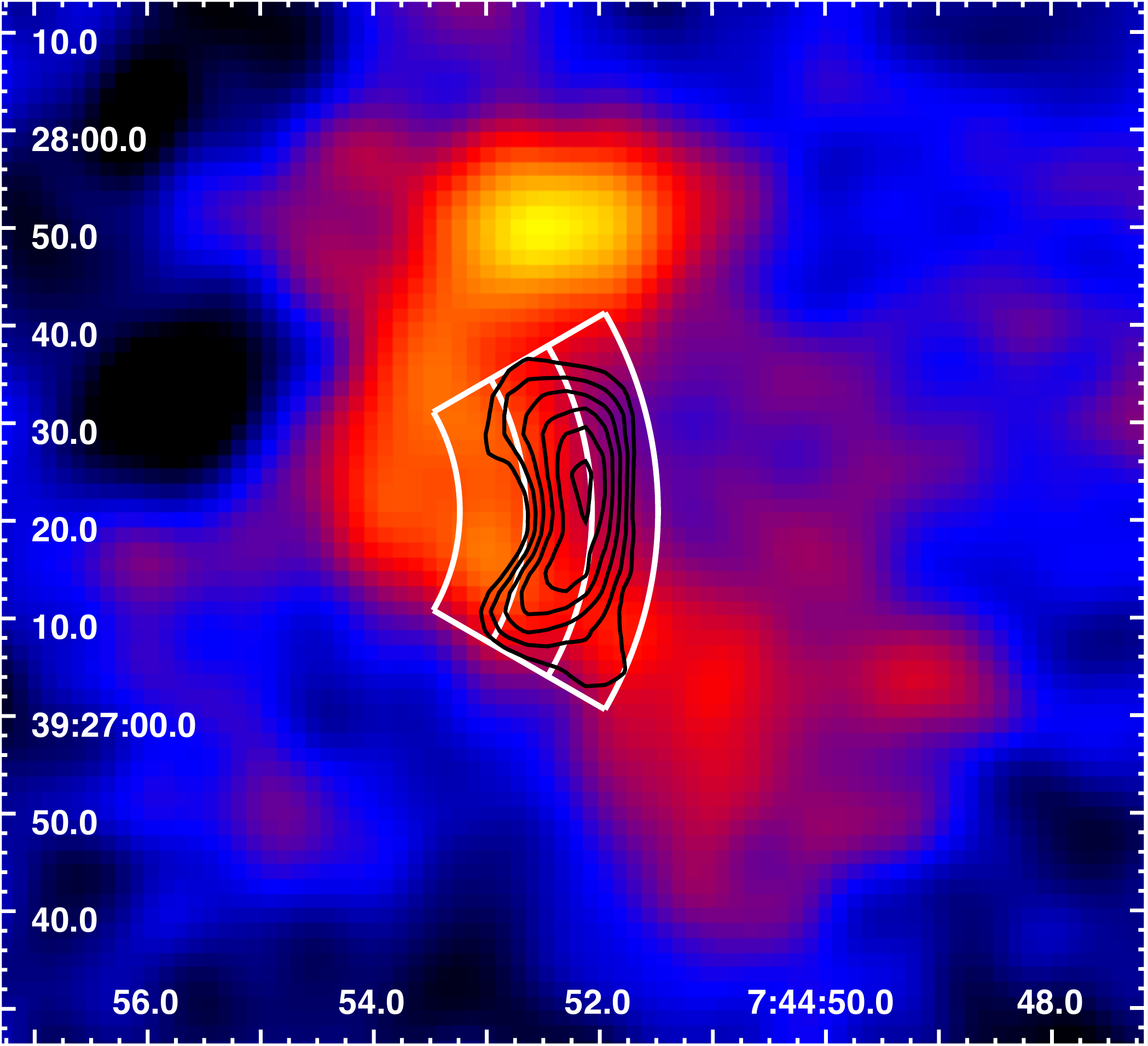}
\caption{Pie cut region (shown in white) in which the flux density is measured to determine a radio relic upper limit. Black contours are from the MUSTANG map. The radio relic upper limit is measured from our compact-source-subtracted LOFAR image with an outer $uv$-taper of 6 arcsec. \label{relreg}}
\end{figure}

\begin{figure}
\centering
\includegraphics[width=0.49\textwidth]{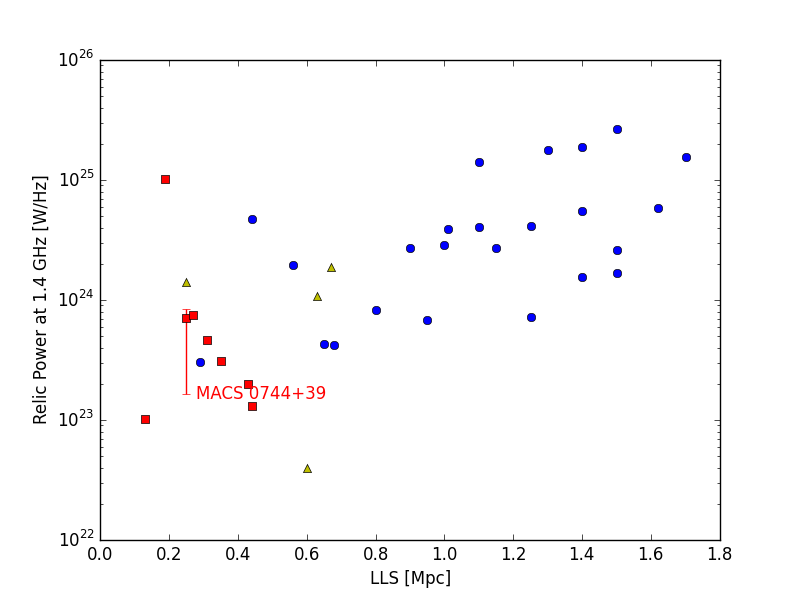}
\caption{A sample of gischt radio relics (blue circles) and AGN/phoenix relics (red squares) from \citet{2017MNRAS.470..240N} plotted by their power at 1.4 GHz versus their LLS in Mpc. Yellow triangles represent additional relics / candidate relics from Table~\ref{tab1}. The red line represents the upper limit range of the power for the radio emission detected within the shocked region for MACS J0744.9+3927. Powers include a k-correction to account for redshift, with an average spectral index of $\alpha = -1.2$, as is used in \citet{2017MNRAS.470..240N}. \label{plot}}
\end{figure}

There is no clear radio structure coinciding with the shock detected by SZE observations. The only emission present at the shocked region is what we suspect to be the fading southwestern edge of the AGN lobe associated with the BCG. To determine an upper limit on a potential radio relic associated with the shock, we calculate the radio power within the shocked region from our low-resolution image (14 arcsec $\times$ 12 arcsec) made with the compact-source-subtracted $uv$ dataset that was imaged in CASA {\tt CLEAN} with an outer $uv$taper of 6 arcsec (Fig.~\ref{t6}). \\

The flux density at 143 MHz is measured within a pie cut of three annuli, where the beginning and ending angles of the pie align with the north and south boundaries of the lowest contour (3$\sigma$) of the SZ decrement as seen by MUSTANG (presented in \citealp{2011ApJ...734...10K}). Fig.~\ref{relreg} shows the pie cut annuli region in white and the SZ ridge in black overlaid on our low-resolution LOFAR image. The outer two annuli effectively cover the area of the shock as indicated by SZE observations, and the inner annulus extends from the inner edge of the kidney-shaped ridge toward the cluster centre\footnote{The cluster centre as marked by the brightness distribution in X-ray emission.}.
\\

Within a region defined by the outer two annuli shown in Fig.~\ref{relreg}, the flux density at 143 MHz is $S_{143} = 1.18 \pm 0.12$ mJy with a corresponding radio power of $P_{143} = (2.84\pm0.28)\times 10^{24}$ W Hz$^{-1}$. Within a region defined by the total pie cut, including the inner annulus, the flux density is found to be $S_{143} = 2.05 \pm 0.21 $ mJy with a corresponding radio power of $P_{143} = (4.90\pm0.49)\times 10^{24}$ W Hz$^{-1}$. Here we include a k-correction to account for redshift using a spectral index of $\alpha = -1.2$ where $S \propto \nu^{\alpha}$. The error in our flux density measurement is assumed to be 10\%, which is determined by comparing flux densities of several sources in our LOFAR map to the same sources in TGSS \citep{2017A&A...598A..78I} and the 7C survey \citep{2007MNRAS.382.1639H}.\\

We extrapolate the flux density value within the total pie cut, defined by the three annuli, to 1.4 GHz by assuming a range of potential spectral indices that include the typical values measured for relics found at merger shocks (gischt-type relics) and AGN-relics which can have even steeper spectral indices. We choose a conservative range of $\alpha= -1$ to $-1.5$. This gives a 1.4 GHz flux density range of $S_{1.4} =(0.06-0.23)$ mJy, corresponding to a radio power range of $P_{1.4} = (1.71-4.99)\times 10^{23}$ W Hz$^{-1}$.\\

In the above calculations for flux density and radio power at 1.4 GHz, we choose to use the 143 MHz flux density within the total pie cut (which is a region starting at the cluster centre and ending at the outer edge of kidney-shaped ridge, $\sim 140$ kpc in width) rather than just using the flux which is coincident with the MUSTANG SZE contours. An ALMA detection from \citet{2016ApJ...829L..23B} showed that a pressure discontinuity (SZ decrement) of the merger shock in El Gordo is coincident with the width of the radio relic seen at 2.1 GHz. However, MUSTANG has a lower resolution and lower sensitivity than ALMA, and radio relics are seen to widen at low frequencies. For this reason, we calculate the potential relic flux density from annuli covering the shocked region as indicated by SZE observations in addition to the flux within an annulus extending east of the SZ ridge toward the cluster centre. \\

In Table~\ref{tab1} we make a list of observed relics / candidate relics that are smaller (largest linear scale (LLS) $< 0.7$ Mpc) and/or fainter (log$_{10}$(P$_{1.4}$) $<$ 24.3 W Hz$^{-1}$)\footnote{We note that the small relic in MAXBCG138+25 is perhaps an important outlier since it is much brighter than other small phoenix/AGN relics.} than most gischt relics. In Fig.~\ref{plot} we plot a large sample of observed radio relics by their power at 1.4 GHz versus their LLS. This sample is taken from a list of confirmed gischt-type relics and smaller phoenix relics presented in \citet{2017MNRAS.470..240N} as well as the additional smaller relics / candidate relics listed in Table~\ref{tab1}. We calculate the power for each relic from the flux density at 1.4 GHz and include a k-correction with an average spectral index of $\alpha = -1.2$, as is used in \citet{2017MNRAS.470..240N}. We include the extrapolated radio power range for the shocked region in MACS J0744.9+3927, using the LLS of the shock as seen by MUSTANG (250 kpc). The largest value of this power range is fainter than that of most observed gischt-type relics (shown as blue circles in Fig.~\ref{plot}), however, since the LLS is smaller than most gischt-type relics, a lower power would be expected. On the basis of the measured power range alone, a gischt-type relic cannot be ruled out. The power range and LLS agree more closely with powers and sizes seen in phoenix-type relics (shown as red squares in Fig.~\ref{plot}). \\

Based on the location of the shock and the LLS, it does not seem likely that a gischt-type relic would be generated, and indeed we see no such clear radio structure resembling a gischt relic at the shock site. The absence of a morphological structure resembling a relic is interesting, particularly in light of the high electron acceleration efficiency observed in radio relics, which is still poorly understood \citep{2014MNRAS.437.2291V}. 

\subsection{Upper limit on particle acceleration efficiency}
Taking the parameters of the shock wave detected in MACS J0744.9+3927 (an average Mach number of $\mathcal{M} = 1.75$ and shock velocity $V_{\rm sh}=1827$ km s$^{-1}$) and the non-detection of a radio relic, we can compute an upper limit on the particle acceleration efficiency. Comparing the dissipated kinetic power at the shock to the total power in the radio emission, we can estimate the acceleration efficiency using equation 2 in \citet{2016MNRAS.460L..84B}:

\begin{equation}
\int_{\nu_{0}} L(\nu) d\nu 	\simeq \frac{1}{2} \eta_{\rm e} \Psi \rho_{\rm u} V_{\rm sh}^{3} (1 - C^{-2}) \frac{B^{2}}{B_{\rm cmb}^{2} + B^{2}} S ,
\end{equation}
where $\eta_{\rm e}$ is the acceleration efficiency, $\rho_{\rm u}$ is the upstream density, $V_{\rm sh}$ is the shock velocity, $C$ is the compression factor which is related to the Mach number via $C = 4 \mathcal{M}^{2} / (\mathcal{M}^{2} + 3)$, $B$ is the magnetic field strength and $B_{\rm cmb} = 3.25(1+z)^2$, $S$ is the surface area of the shock\footnote{This area is the largest linear length times the largest linear width of the shocked region, defined by the pie cut in Fig.~\ref{relreg} ($250 \times 140$ kpc$^{2}$).}, and $\Psi$ is the ratio of the energy injected in electrons emitting over the full spectrum versus electrons emitting in radio wavelengths, given by

\begin{equation}
\Psi = \frac{\int_{p_{0}} Q(p) E(p) dp}{\int_{p_{\rm min}} Q(p) E(p) dp} ,
\end{equation}
where $Q(p) \propto p^{-\delta_{\rm inj}}$ and $\delta_{\rm inj} =  2(\mathcal{M}^{2} + 1)/(\mathcal{M}^{2} −1)$ \citep{1987PhR...154....1B}. The momentum, $p_0$, is the momentum associated with electrons that emit the characteristic frequency of the synchrotron emission, $\nu_0=p_0^2 e B /2\pi m_{\rm e}^3c^3$. Here, $m_{\rm e}$ is the electron mass, $e$ its charge, and $c$ the speed of light. For the minimum momentum in the denominator, $p_{\rm min}$, we consider two cases: 1) for a low value of $p_{\rm min} = 0.1 m_{\rm e}c$ the efficiency has to be unrealistically high, $\gg 100\%$, and we cannot infer an upper bound for $\eta_{\rm e}$, or 2) the shock re-accelerates a population of relativistic seed electrons with $p_{\rm min} = 100 m_{e}c$, in which case the efficiency must be $\leq 19\%$ since a relic is not observed. Clearly, this number is uncertain, since it depends not only on $p_{\rm min}$, but also quite strongly on the assumed velocity of the shock, $V_{\rm sh}$, the upstream density, $\rho_{\rm u}$, as well as the magnetic field, $B$. We used the value from Figure 5 in \citet{2011ApJ...734...10K} for the upstream density ($\rho_{\rm u} = 0.013~$cm$^{-3}$), and since the magnetic field in this cluster is not known we assume a value of $B=1~\mu$G.

\subsection{Missing radio halo}\label{3p4}
The study of radio halos in high-mass and high-$z$ merging clusters is crucial to constrain the model of turbulent re-acceleration. In the turbulent re-acceleration scenario, it is expected that the fraction of ultra-steep-spectrum radio halos increases strongly with redshift because of stronger inverse Compton losses \citep{2010A&A...509A..68C}. \\

\begin{figure}
\centering
\includegraphics[width=0.49\textwidth]{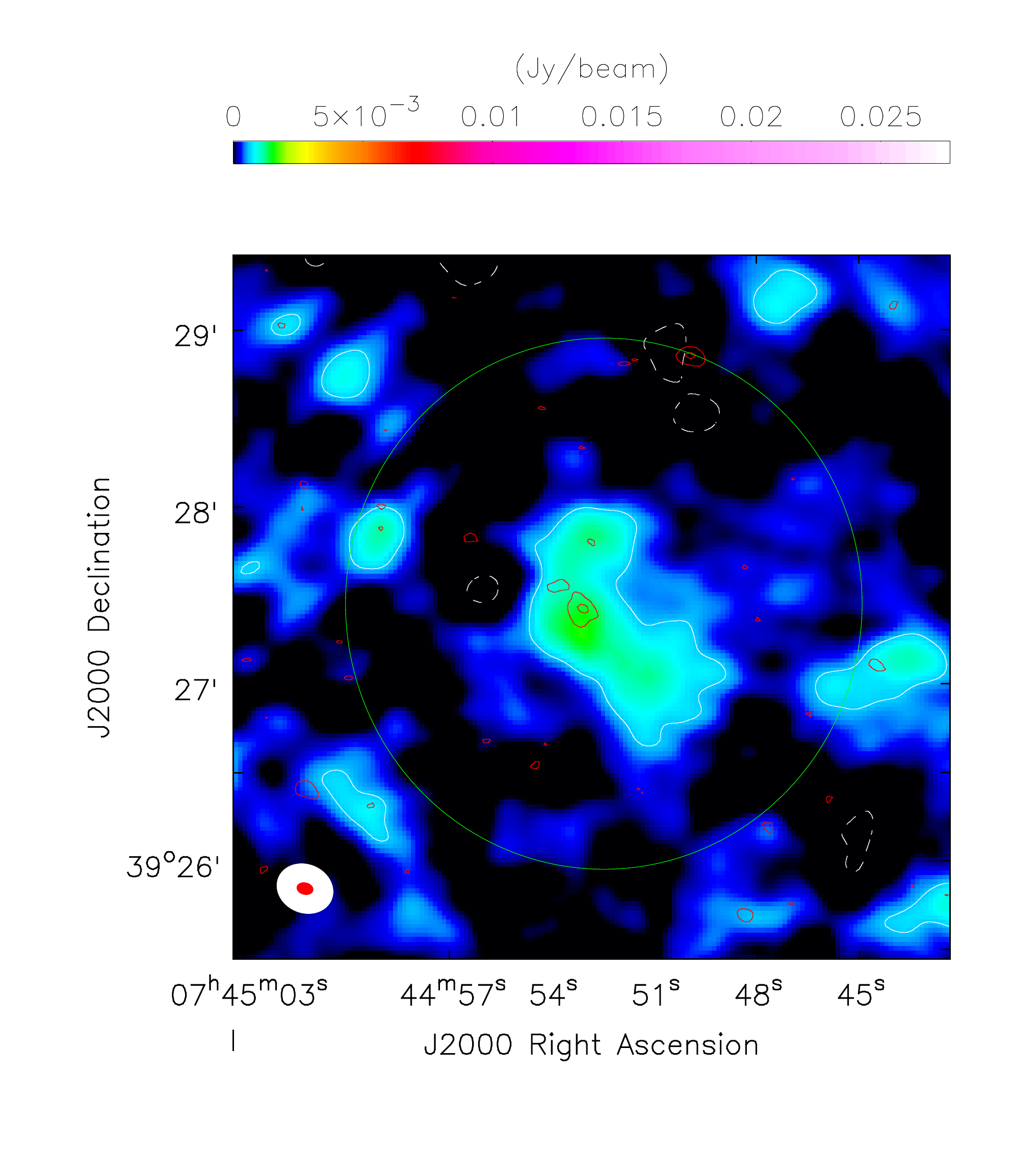}
\caption{LOFAR image after subtracting sources imaged at a $uv$-range of $>$ 4000 $\lambda$. An outer $uv$taper of 10 arcsec was used to bring out diffuse emission. RMS noise is $\sigma=$200$\mu$Jybeam$^{-1}$. White contours represent $[-3,3]~\times~\sigma$. Red contours $[3,12]~\times~\sigma$ are the compact sources imaged with $uv$-range $>$ 4000 $\lambda$, which were subtracted. The green circle represents the area that would be expected for a radio halo in a cluster of this mass, with a radius of 650 kpc. We use flux density measurements in this region to determine a the radio halo power upper limit. \label{halo}}
\end{figure}

\begin{figure}
\centering
\includegraphics[width=0.49\textwidth]{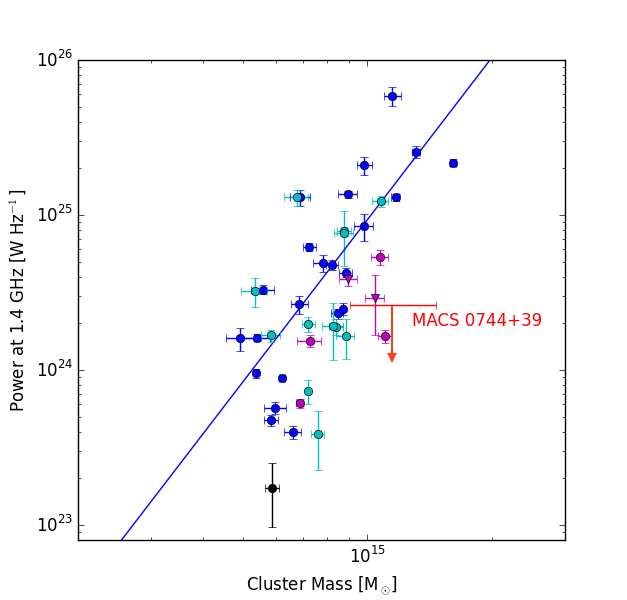}
\caption{A sample of radio halos plotted by their radio power at 1.4 GHz versus their cluster mass (M$_{500}$ -- as determined from Planck observations). The sample of halos and their correlation is reproduced from \citet{2016A&A...595A.116M}. Halos with flux measured at 1.4 GHz are marked by blue circles and their derived fit is shown as a blue line. Cyan circles represent halos with flux measured at frequencies other than 1.4 GHz. Magenta circles represent ultra-steep halos, and magenta triangles represent ultra-steep halos with flux measured at frequencies other than 1.4 GHz. The ultra-steep-spectrum radio halo in Abell 1132 is also included from \citealt{2017arXiv170808928W}, and is marked by a black circle. The upper limit of radio halo power at 1.4 GHz in MACS J0744.9+3927 is represented by the red arrow. We note that this power is likely an overestimation since it consists of AGN lobe power. All the halo powers include a k-correction with an averaged spectral index of $\alpha = -1.3$, as in \citet{2016A&A...595A.116M}. \label{corr}}
\end{figure} 

In MACS J0744.9+3927, the LOFAR image does not show a radio halo, even though the X-ray and SZ data indicate that this system is in the process of a merger, albeit not a major one. Diffuse emission at the cluster centre is most likely caused by the active BCG as well as the galaxy north of it whose redshift is unknown. If there is radio halo emission, it is eclipsed by the emission of the active galaxies, whose lobes extend to $\sim 570$ kpc. However, a high-mass cluster such as this would be expected to host a radio halo on larger scales \citep{2007MNRAS.378.1565C}. \\

To determine an upper limit on the flux of a radio halo, we use an image made after subtracting compact sources imaged at a $uv$-range of $>$ 4000 $\lambda$ (corresponding to emission spanning less than $\sim 400$ kpc) with rms noise of $\sigma = 200~\mu$Jy beam$^{-1}$ (see Fig.~\ref{halo}). We estimate the upper limit of a radio halo by defining a circular region (shown as the green circle in Fig.~\ref{halo}) with an origin at the cluster centre and a radius of 650 kpc, as would be expected for a cluster this mass. The upper limit on the flux density is expressed as the summation of flux density from two regions within this circle: 1) the flux density within 3$\sigma$ contours of the central diffuse emission (which includes AGN emission), and 2) the rms noise, $\sigma$, times the number of beams covering the remaining area inside the circle and outside of the central 3$\sigma$ contours. We state that this upper limit of $S_{143} = 19.9~$mJy is probably an overestimate since it clearly includes emission originating from AGN. Assuming a spectral index of $\alpha = -1.3$ and including k-correction, we find that the extrapolated power at 1.4 GHz of this upper limit is $P_{1.4} = 2.61 \times 10^{24}$ W Hz$^{-1}$ and falls below the correlation for radio halo power versus cluster mass for a sample of clusters (see Fig.~\ref{corr}). Since there are no Planck observations of this cluster, we must use the cluster mass as derived by weak-lensing from CLASH\footnote{The CLASH mass range is consistent within $1\sigma$ with the cluster mass value derived from XMM Newton X-ray data.}. Given the large errors in the mass for MACS J0744.9+3927, the upper limit for the radio power shown in Fig.~\ref{corr} is not far from the correlation at the lower end of the mass estimate. Moreover, for a given mass, the scatter in radio halo power is quite large, and the correlation may change once deeper and more systematic searches for halos are underway.\\

\begin{figure}
\centering
\includegraphics[width=0.49\textwidth]{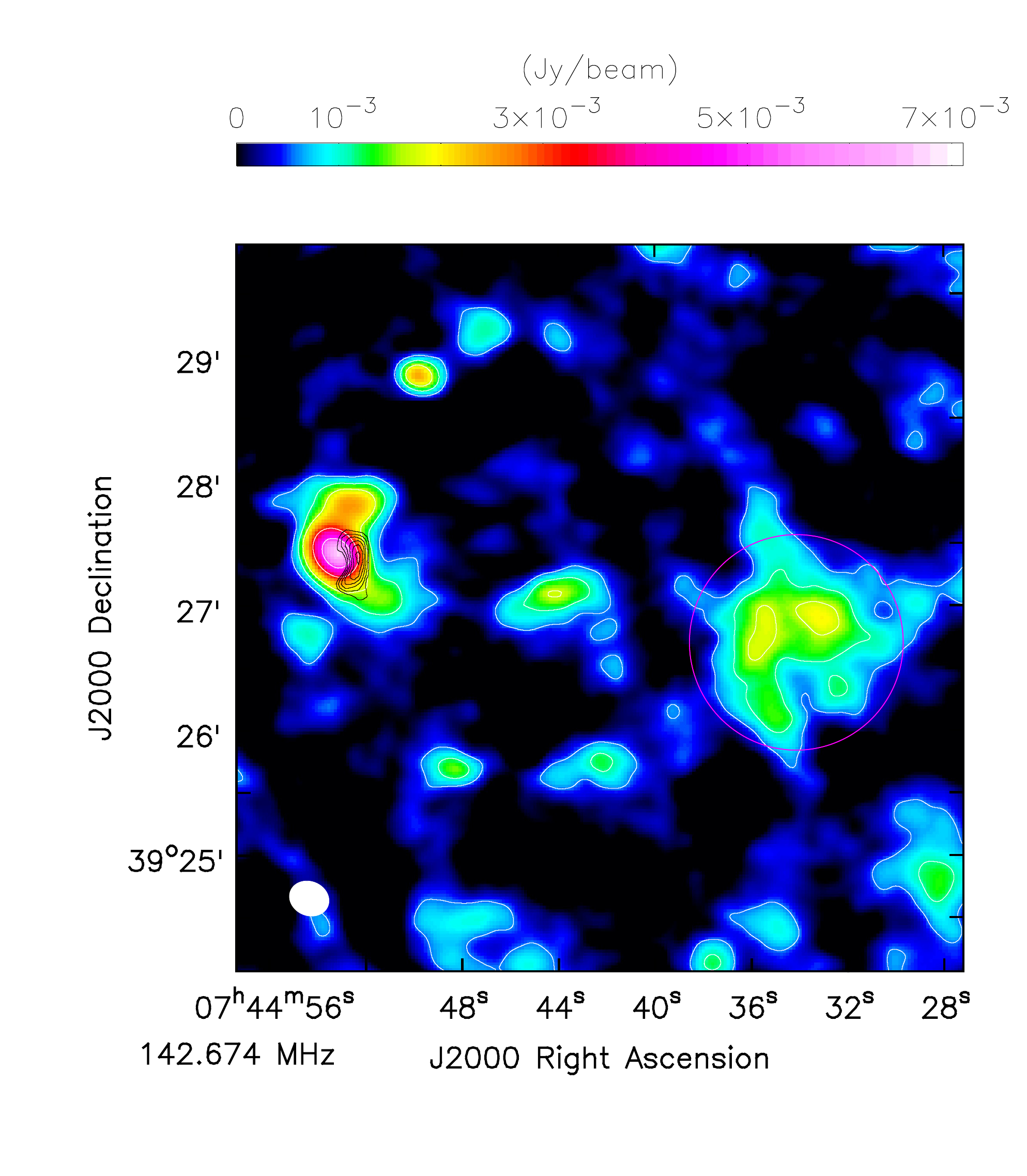}
\caption{LOFAR image after injecting a simulated radio halo, with a power of $P_{1.4} = 8.54 \times 10^{24}$ W Hz$^{-1}$ at 1.4 GHz and a spectral index of $\alpha = -1.5$, to the west of the cluster centre. The magenta circle has a radius of $3r_e$ centered on the coordinates where the simulated halo was injected. White contours are $[2, 4, 6, 12, 18]~\times~\sigma$ where $\sigma=300~\mu$Jybeam$^{-1}$. The flux density of the recovered halo is measured within the circular magenta region, above 2$\sigma$, and is found to be $20.02 \pm 2.00~$mJy. \label{mockhalo}}
\end{figure}

\subsection{Radio halo injection}
We simulate a radio halo by injecting a radio source into the $uv$ data that has a central brightness $I_{\rm 0}$ and an e-folding radius of $r_{e}$, which is defined as the radius at which the brightness drops to $I_{\rm 0}/e$ \citep[e.g.][]{2007ApJ...670L...5B, 2008A&A...484..327V,2017MNRAS.470.3465B}. The length scale $r_{e}$ is therefore relatively independent on the sensitivity of the radio images. The radio halo we inject has a size and brightness specified by the correlation in Fig.~\ref{corr} ($P_{1.4} = 8.54 \times 10^{24}$ W Hz$^{-1}$ and $R_{H} = 650$ kpc for a cluster with $M_{500} = 9.9\times10^{14}M_{\odot}$), such that $I_{\rm 0} = 0.115~\mu$Jy arcsec$^2$ and $r_e$ = 250 kpc. The model of this mock radio halo is Fourier transformed into the visibility data (MODEL\_DATA column), taking into account the w-projection parameter. A relatively empty region near the cluster centre, void of bright sources or artefacts, is chosen to host the injected flux (at RA: 07h44m34.12s, Dec: +39$^{\circ}$26\arcmin42.3\arcsec). The dataset is then re-imaged with Briggs' robust parameter 0 and an outer $uv$-taper of 10 arcsec. Since we set the power at 1.4 GHz, a flatter spectral index would translate to a weaker power at 143 MHz than a steeper spectral index\footnote{Radio halos are usually shown to exhibit spectral indices < -1.}. We adjust the spectral index from $\alpha = -1.0$ to $-1.8$ and determine when the injected flux is visible at 143 MHz. The injected halo is considered detected when it is recovered above 2$\sigma$ with a diameter of roughly $3r_{e}$. We find that the spectral index must be $\alpha \leq -1.5$ for the halo to be recovered in our LOFAR images. With a spectral index of $\alpha = -1.5$, the total integrated flux density of the injected halo is 26.6 mJy at 143 MHz. Our LOFAR image of this simulated halo is shown in Fig.~\ref{mockhalo}. \\

The flux density of the recovered halo is then measured in our LOFAR image within a region centered on the coordinates of the injected halo with a diameter of roughly $3r_{e}$. The integrated flux density of the recovered halo, above 2$\sigma$, is found to be $20.02 \pm 2.00~$mJy. This observed flux density is 75\% that of the injected flux density. Compared to our upper limit determined in Sec.~\ref{3p4}, the recovered flux density of the mock halo is approximately equal. If a halo with a spectral index $\leq -1.5$ was present in this cluster, it is likely that our flux density upper limit would have been greater than the flux density of the recovered mock halo since our upper limit was measured including AGN emission as well. We argue that a halo with spectral index $\leq -1.5$ is not present in this cluster, and that a halo with a flatter spectral index may exist but cannot be detected by our LOFAR observations at 143 MHz.

\section{Discussion and Conclusion}

Sunyaev-Zel'dovich and X-ray observations have revealed a shock, with Mach number $\mathcal{M} = 1.0-2.9$ and a length of $\sim 200$ kpc, near the centre of the cluster MACS J0744.9+3927. To search for diffuse radio emission associated with the merger, we have imaged the cluster with LOFAR at 120-165 MHz. Our LOFAR radio images do not show a radio relic coincident with the shock nor the presence of a radio halo at the cluster centre. With its estimated Mach number close to the Mach numbers of giant shock waves observed on cluster outskirts (which form powerful radio relics such as the Toothbrush relic) a search for radio emission associated with the shock in MACS J0744.9+3927 is important to understand the mechanisms of particle acceleration in the ICM.\\

Although the shock detected by MUSTANG is considered to be merger-induced by \citet{2011ApJ...734...10K}, it is very different from shocks that produce gischt-type relics, which are typically found on the cluster outskirts, since it is smaller and near the cluster centre. Instead of being induced by a merger, the shock may have been caused by an outburst of the central AGN. However, in \citep{2011ApJ...734...10K} it is also reported that there is the presence of a cold front behind the shock; this lends support to the merger-induced scenario for the shock front. Interestingly, simulations show bright mock relics occurring close to the cluster centre \citep{2017MNRAS.470..240N}; however, these have not yet been confirmed by observations. \\

The upper limit of the radio luminosity in the shocked region also suggests that no DSA at the shock front takes place as this process generally leads to higher luminosities relative to the size of the source (see Fig.~\ref{plot}). It is unclear why no DSA takes place, at least not with the efficiency that is observed in large radio relics. The magnetic field direction may have an influence on the efficiency of electron acceleration \citep[e.g.][]{2014ApJ...797...47G}. However, \citet{2017MNRAS.464.4448W} have shown that the magnetic field distributions in galaxy clusters as predicted by cosmological magneto-hydrodynamical simulations have little effect on the radio luminosities of radio relics.\\

If pre-existing populations of older cosmic ray (CR) electrons are required for the injection into a DSA process, one would expect that close to a central radio galaxy there should be no shortage of seed CR-electrons: a phoenix-type relic caused by the re-acceleration of old AGN lobes might be expected in MACS J0744.9+3927. The AGN should contribute mildly energetic electrons which could then be re-accelerated via compression by the shock front. Another recently proposed mechanism is ``gentle reenergization" \citep{2017arXiv171006796D}, in which old CR-electrons get accelerated on time-scales larger than sound crossing times to produce steep and filamentary radio emission, but this does not appear to happening in MACS J0744.9+3927. \\

While we suggest that the radio flux within the shocked region is probably not attributed to a gischt-type relic, due to the small size and central location of the shock,
deeper upper limits on the radio power would be required to rule out this possibility. The upper limit of the radio power and the LLS of the shocked region agree more closely  to known phoenix/AGN relics, but we argue that the emission present is not actually re-energized relic emission but simply a contamination from the original AGN lobe emission of the BCG. \\

Since we do not see a relic, or a re-brightening of AGN emission (in the form of a phoenix), it could be the case that the shock and the AGN are not in the same plane and that there is not a sufficient supply of seed electrons being injected into the shocked region. We find that the energy dissipated at the shock would be insufficient to accelerate a population of only very mildly relativistic electrons ($p_{\rm min}=0.1 m_{\rm e}c$). \\

There is some disparity between the Mach numbers as determined from X-ray observations versus radio observations. It has been shown that radio emission of relics traces higher Mach numbers than those inferred from temperature or brightness discontinuities in X-ray \citep{2015ApJ...812...49H, 2015PASJ...67..113I, 2015A&A...575A..45T, 2017A&A...600A.100A}. However, a Mach number of $\mathcal{M} = 2.1^{+0.8}_{-0.5}$ as inferred from MUSTANG and Chandra data is comparable to Mach numbers found, e.g. in the Toothbrush relic.\\

Finally, we also see no signs of a giant radio halo despite that this cluster is massive ($M_{500} = (11.837 \pm 2.786 )\times10^{14}M_{\odot}$; \citealt{2015MNRAS.449.2024S}) and considered to be a merger. Radio halos at high redshifts are expected to have shorter life times because of the higher inverse Compton losses. Hence, one would expect a higher fraction of ultra-steep spectrum halos compared to lower redshifts. Still, the El Gordo cluster, which is one of the highest redshift ($z=0.9$) and most massive merging clusters known, exhibits, both, radio relics and a radio halo that can be seen over a range of radio frequencies. Since MACS J0744.9+3927 has a mass similar to that of the El Gordo cluster, and is at a slightly lower redshift, it is surprising that a radio halo is not visible at low radio frequencies.\\

A simulation of a radio halo injected into our LOFAR data proves that a halo falling on the correlation line for a cluster this mass, with a spectral index $\alpha \leq -1.5$, is not present in the cluster. This cluster may host a radio halo with a flatter spectrum to which our low-frequency observations are not sensitive. This also brings into question the merger phase of this cluster. If it is in an early phase, it may explain why the merger shock is small and so close to the cluster center, and also why a radio halo with steep spectrum emission is not yet visible. 

\section*{Acknowledgements}

This work was supported by the Deutsche Forschungsgemeinschaft (DFG) through the Collaborative Research Centre SFB 676 ``Particles, Strings and the Early Universe", project C2. LOFAR, the Low Frequency Array designed and constructed by ASTRON, has facilities in several countries, that are owned by various parties (each with their own funding sources), and that are collectively operated by the International LOFAR Telescope (ILT) foundation under a joint scientific policy. The LOFAR software and dedicated reduction packages on \url{https://github.com/apmechev/GRID_LRT} were deployed on the e-infrastructure by the LOFAR e-infragroup, consisting of J. B. R. Oonk (ASTRON \& Leiden Observatory), A. P. Mechev (Leiden Observatory) and T. Shimwell (Leiden Observatory) with support from N. Danezi (SURFsara) and C. Schrijvers (SURFsara). This work has made use of the Dutch national e-infrastructure with the support of SURF Cooperative through grant e-infra160022. This work has made use of the Lofar Solution Tool (LoSoTo), developed by F. de Gasperin. This research has made use of the NASA/IPAC Extragalactic Data Base (NED) which is operated by the JPL, California institute of technology under contract with the National Aeronautics and  Space administration. AB acknowledges supports from the ERC Stg 714245 DRANOEL. RvW acknowledges support from the ERC Advanced Investigator programme NewClusters 321271. DW acknowledges support by the DFG through grants SFB 676 and BR 2026/17 and by the ERC through Project No. 714196.




\bibliographystyle{mnras}
\bibliography{MacsJ0744} 







\bsp	
\label{lastpage}
\end{document}